\documentclass{article}

\usepackage{PRIMEarxiv}

\usepackage[utf8]{inputenc} 
\usepackage[T1]{fontenc}    
\usepackage{hyperref}       
\usepackage{url}            
\usepackage{booktabs}       
\usepackage{amsfonts}       
\usepackage{nicefrac}       
\usepackage{microtype}      
\usepackage{lipsum}
\usepackage{fancyhdr}       
\usepackage{graphicx}       
\usepackage{bm}
\usepackage{xspace,xcolor}
\graphicspath{{media/}}     

\usepackage{graphicx}
\usepackage{amsmath}
\usepackage{amssymb}
\usepackage{comment}

\newcommand{\methodname}{LyricJam Sonic\xspace}

\title{\methodname: A Generative System for Real-Time Composition and Musical Improvisation
}

\author{
  Olga Vechtomova, Gaurav Sahu \\
  University of Waterloo \\
  Waterloo, ON, Canada\\
  \texttt{\{ovechtom, gsahu\}@uwaterloo.ca} \\
}

\begin{document}
\maketitle

\begin{abstract}
Electronic music artists and sound designers have unique workflow practices that necessitate specialized approaches for developing music information retrieval and creativity support tools. Furthermore, electronic music instruments, such as modular synthesizers, have near-infinite possibilities for sound creation and can be combined to create unique and complex audio paths. The process of discovering interesting sounds is often serendipitous and impossible to replicate. For this reason, many musicians in electronic genres record audio output at all times while they work in the studio. Subsequently, it is difficult for artists to rediscover audio segments that might be suitable for use in their compositions from thousands of hours of recordings. In this paper, we describe \methodname\space-- a novel creative tool for musicians to rediscover their previous recordings, re-contextualize them with other recordings, and create original live music compositions in real-time. A bi-modal AI-driven approach uses generated lyric lines to find matching audio clips from the artist’s past studio recordings, and uses them to generate new lyric lines, which in turn are used to find other clips, thus creating a continuous and evolving stream of music and lyrics. The intent is to keep the artists in a state of creative flow conducive to music creation rather than taking them into an analytical/critical state of deliberately searching for past audio segments. The system can run in either a fully autonomous mode without user input, or in a live performance mode, where the artist plays live music, while the system ``listens'' and creates a continuous stream of music and lyrics in response.\footnote{Demo video: \url{https://vimeo.com/755240824/4345443ab6}}.

\keywords{Lyrics Generation  \and Generative Music \and Neural Network \and Variational Autoencoder \and Generative Adversarial Network.}
\end{abstract}
\section{Introduction}
Electronic music artists and sound designers can create complex and unique audio paths by patching multiple modules and standalone instruments to achieve a desired sound effect. Such audio paths typically include a combination of analog and digital sound processing. 
The process is inherently volatile, resulting in sounds that can be impossible to replicate, especially in complex audio paths. For this reason, electronic artists often record all of their studio sessions. 

Another aspect of electronic music that sets it apart from other genres is a composition style that electronic instruments, such as modular synthesizers, afford. While it is certainly possible to approach electronic music composition by writing a score and playing it on a synthesizer keyboard, electronic artists often have an organic approach to composition, where they may start with an open mind and get inspiration by playing the instruments~\footnote{\url{https://www.factmag.com/2020/01/08/alessandro-cortini-in-the-studio/}}. This is in part due to the experimental nature of the instruments and relative importance of non-musical sound effects and textures (e.g, noise and drone) in electronic and electro-acoustic compositions, especially in such genres as ambient. Due to these distinctive aspects of electronic music and its composition, artists can accumulate thousands of hours of studio recordings. Sifting through and listening to hours of recordings is impractical and can take the artists out of their creative flow. In this work, we envisioned and implemented a creative tool, \textbf{\methodname}, for artists to: 1) tap into their catalogue of studio recordings, 2) rediscover sounds, 3) re-contextualize them with other sounds and 4) create new original music compositions or soundscapes using this tool.

An important consideration that such tool must have is to be conducive to creativity and not take the artist out of their creative flow. For this reason, our tool was envisioned as facilitating serendipitous discovery, rather than enabling a deliberate search process controlled by the artist. The tool is therefore close to the philosophy of serendipitous discovery in electronic music composition.
\methodname embodies two neural networks that interact with each other and generate continuous and evolving stream of music and lyric lines. One neural network generates lyric lines, while the other uses the generated line to find a congruent audio clip from the artist’s catalogue of studio recordings. The audio clip is integrated into the continuous music stream played to the artist and is used in turn to generate another line, which continues the process.

Past research has shown that conditional lyrics generation based on audio clips of instrumental music results in lyrics that are emotionally congruent with the music they are conditioned upon~\cite{vechtomovaLyrics2020,vechtomovalyricjam}. We expect that emotional valence of lyrics will similarly lead to congruent audio clips. The process can be initiated with an artist supplying a starting line to set the emotional tone, but it can also be initiated from a randomly generated line if the artist does not want to take the system in a specific direction. 

While the system can run fully autonomously without the artist intervention, the artist can override the system at any time by providing their own lyric lines or by playing live music to the system. In the live performance mode, the user plays musical instruments, while the system ``listens'' and creates a continuous stream of music based on the artist's own past audio clips, and generates corresponding lyrics.

The process of interaction between two neural networks results in a smooth and coherent generative musical composition, as opposed to incoherent audio clips spliced together. The process re-contextualizes audio segments from original recordings into a new composition, which not only lets the artists hear audio segments that they may otherwise never hear, but also inspire them to combine these segments in new ways. Furthermore, the system can be used by an artist to autonomously generate new original generative compositions from the artist's catalogue of studio recordings. The rationale for using lyrics as a vehicle for finding audio clips, as opposed to just using the previously played clip to find another one, is to introduce progression in the generative music, and also bring about elements of surprise. If the next sound clip was selected based solely on its degree of similarity to the previous clip, it would lead to monotonous compositions, effectively preventing any kind of musical transition and development. Lyric generation is not a goal in itself for this tool but an essential step in the music generation; however, generated lyric narrative can also be used for inspiration if an artist so desires.

\section{Related Work}
The present work builds upon LyricJam \cite{vechtomovalyricjam}, a system for generating a stream of lyrics in real-time based on the audio of music, played live by a musician. While LyricJam was envisioned as a tool for inspiring the musician during live performance, the main goal of the present system is to help an artist get inspiration not only from generated lyric lines, but also help them discover congruent sounds from their catalogue of studio recordings, and seek inspiration from possibly unexpected sequences of sound clips suggested by the system. 

There is a large body of work in the area of generative music \cite{Hunt2017ThoughtsOI,briot2019deep}, which predominantly includes approaches to generating original raw audio \cite{oord2016wavenet} or music score \cite{DUA2020465,lstmmusicgeneration}. A number of approaches address generation of specific instrumental accompaniment to existing music, e.g. drum tracks \cite{conditional_drums}; however, this body of work is largely not relevant to this paper as the proposed system does not generate raw audio or music score, but creates an original music stream from existing clips in the artist's catalogue. 

Some of the earliest approaches to generative music that construct a new musical composition by remixing existing sound sources were pioneered in the 60s and 70s by electronic musicians. Notably, Brian Eno \cite{BrianEno} created generative music by capturing the output of multiple magnetic tape recorders running at different speeds, each playing a single note. More recent algorithmic approaches to creating generative soundscape compositions through sampling or retrieving clips of existing recordings include \cite{eigenfeldt2011negotiated,thorogood2012audio,Turchet2020VoicebasedIF}, which combine heuristic rules with ranking functions to retrieve the next clip.

Most of the work in generative music domain is aimed at developing tools for creating new compositions. In contrast, in our work, while the artist may choose to use the resulting music or its parts as a composition in its own right, the intent is to help the artist find new inspiration based on their own past recording sessions.


The main contribution of this work is a bi-modal self-perpetuating system, where a lyric line is generated in real-time based on the previously played clip, both of which influence the selection of the next clip. The system is based on neural networks, which are trained to predict the next clip based on the intrinsic properties of the raw audio of the previous clip and its accompanying lyric line. Furthermore, the system is trained to generate lyric lines conditioned on the raw audio, which differs from past works on lyric generation primarily conditioned on rules, e.g. melody and style~\cite{seqgan-melodyconditioned,malmi2016dopelearning,oliveira2015tra,potash2015ghostwriter,watanabe2018melody}.

\section{Background}
\subsection{Variational Autoencoder (VAE)}
\label{sec:bg}
A variational autoencoder (VAE)~\cite{kingma2014auto} is a stochastic neural generative model that consists of an encoder-decoder architecture.
The key idea is to learn a posterior distribution that closely approximates the true posterior (also called prior).
The distance between the approximate posterior and the prior is minimized using KL divergence, a distance metric in the probabilistic space.
The VAE objective minimizes the evidence lower bound (ELBO) given by the following expression:
\begin{align}
\label{eq:elbo}
    ELBO = E_{z \sim q_\phi}[\log p_\theta(x | \bm z)] - KL(q_\phi(\bm z | x) || p_\theta (\bm z)),
\end{align}
where $x$ is the input, $\bm z$ is a latent variable, $q_\phi(\bm z|x)$ is the posterior distribution to learn, $p_\theta (\bm z)$ is the true prior, and $KL(q_\phi(\bm z| x) || p_\theta(\bm z)) = E_{z \sim q_\phi}[\log q_\phi(\bm z | x) - \log p_\theta (\bm z))]$.
Notably, the parameters $\phi$ and $\theta$ are jointly learned.
VAE improves over a vanilla autoencoder by learning a continuous latent space as opposed to a deterministic input-output mapping; this allows us to sample diverse outputs from the model at inference time.

\subsection{Generative Adversarial Network (GAN)}
A generative adversarial network (GAN)~\cite{goodfellow2014generative} is capable of generating new data samples that match the characteristics of a given data distribution.
It consists of two main components: a generator $G$ and a discriminator $D$.
GAN's objective function incentivizes the generator to fool the discriminator by creating fake samples (also called adversarial samples) that closely resemble the training data distribution, while the discriminator tries to distinguish between an adversarial sample from the generator and a real data sample from the ground truth data distribution.
Formally, GAN optimizes a minimax objective given by the following equation:
\begin{align}
\label{eq:gan_loss}
\min_G \max_D V(D, G) = \mathbb{E}_{x \sim p_{data}(x)} [\log D(x)] + \mathbb{E}_{\bm z \sim p_{\bm z}(\bm z)} [\log (1 - D(G(\bm z)))],
\end{align}
where $x$ is an input sample, $z$ is a latent variable, and $p_{data}(x)$ is the true data distribution we wish the generator $G$ to mimic.

\section{Methodology}
\methodname consists of four main components:
\begin{enumerate}
    \item A Spectrogram Variational Autoencoder \underline{(Spec-VAE)}, which is trained to learn latent representations (latent codes) of the spectrograms corresponding to audio clips.
    \item A Text Conditional Variational Autoencoder \underline{(Text-CVAE)}, which is trained to learn latent representations of lyric lines, conditioned on the the latent code of the corresponding audio clips.
    \item A Generative Adversarial Network \underline{(GAN)} that predicts the latent code of the next audio clip given the element-wise addition of the latent codes of the corresponding lyric line and previous audio clip.
    \item A \underline{Retrieval module}, which retrieves the audio clip from the user's collection based on the cosine similarity of the GAN-predicted latent code and the latent codes of the audio clips in the collection. 
\end{enumerate}

\subsection{Training Spec-VAE}
We train the Spec-VAE model to learn the latent representations of audio clips.
First, we convert the raw waveform audio files into Mel-spectrogram images using the same method as used in \cite{vechtomovaLyrics2020}. These spectrograms are then used as input for the spec-VAE.

The encoder transforms the input spectrogram image $x^{(s)}$ into the approximate posterior distribution $q_\phi(\bm z|x^{(s)})$ learned by optimizing parameters $\phi$ of the encoder\footnote{Superscripts $^{(s)}$ and $^{(t)}$ in our notation refer to spectrogram and text, respectively}. The decoder reconstructs $x$ from the latent variable $\bm z$, sampled from $q_\phi(\bm z|x^{(s)})$. In our implementation, we use convolutional layers as the encoder, deconvolutional layers as the decoder, and a standard normal distribution as the prior distribution $p(\bm z)$.
The VAE loss function combines the reconstruction loss (Binary Cross-Entropy) and KL divergence loss that regularizes the latent space by pulling the posterior distribution to be close to the prior distribution.

\begin{figure*}[!t]
	\centering \small
	\includegraphics[width=1\linewidth]{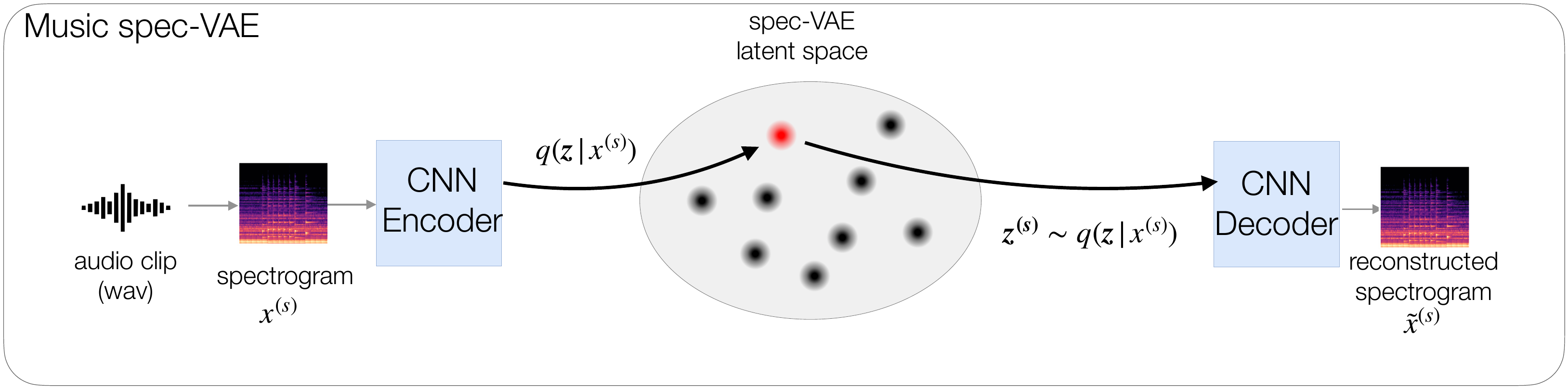}
	\caption{Training spec-VAE.}
	\label{fig:spec-VAE}
\end{figure*}

\subsection{Training Text-CVAE}
Unlike the vanilla VAE used for encoding spectrograms, we use a conditional VAE (CVAE) for encoding lyrics.
The CVAE learns a posterior distribution that is conditioned not only on the input data, but also on a class c: $q_\phi(\bm z|x,c)$. Here, we define the class as the audio clip spectrogram corresponding to a given lyric line. Every conditional posterior distribution $q_\phi({\bm z^{(t)}}|x^{(t}),{\bm z}^{(s)})$ is pulled towards its corresponding prior $p({\bm z^{(t)}}|{\bm z}^{(s)})$. Following prior research on CVAE, all conditional priors are set to the standard normal distribution. 

During training, every input data point consists of a lyric line and its corresponding spectrogram. We first pass the spectrogram through the spec-VAE encoder to get the parameters of the posterior distribution (a vector of means and a vector of standard deviations). We then sample from this posterior to get a vector $\bm {z}^{(s)}$ that is then concatenated with the input of the encoder and the decoder during training. The reason why we used sampling as opposed to the mean ${\bm z}^{(s)}$ vector is to induce the text-CVAE model to learn conditioning on continuous data, as opposed to discrete classes. This prepares it to better handle conditioning on unseen new spectrograms at inference. 
Both the encoder and the decoder in the text-CVAE are LSTMs. The sampled ${\bm z}^{(s)}$ is concatenated with the word embedding input to every step of the encoder and the decoder.
The reconstruction loss is the expected negative log-likelihood (NLL) of data:
\begin{align}
J_\text{rec}(\phi, \theta, {\bm z}^{(s)}, x^{(t)})= -\sum_{t=1}^n \log p(x^{(t)}_i|\bm z^{(t)}, {\bm z}^{(s)}, x^{(t)}_1\cdots x^{(t)}_{i-1})\label{eqn:rec_loss}
\end{align}
where $\phi$ and $\theta$ are parameters of the encoder and decoder, respectively. The overall CVAE loss is given by:
\begin{align}
J = J_\text{rec}(\phi, \theta, {\bm z}^{(s)}, x^{(t)}) + KL(q_\phi(\bm z^{(t)}|x^{(t)}, {\bm z}^{(s)}|p({\bm z}^{(t)}|{\bm z}^{(s)}))
\label{eqn:loss}
\end{align}
where the first term is the reconstruction loss and the second term is the KL-divergence between $\bm z$'s posterior and a prior distribution, which is typically set to standard normal $\mathcal{N}(\bm 0, \mathbf {I})$. 

\begin{figure*}[!t]
	\centering \small
	\includegraphics[width=1\linewidth]{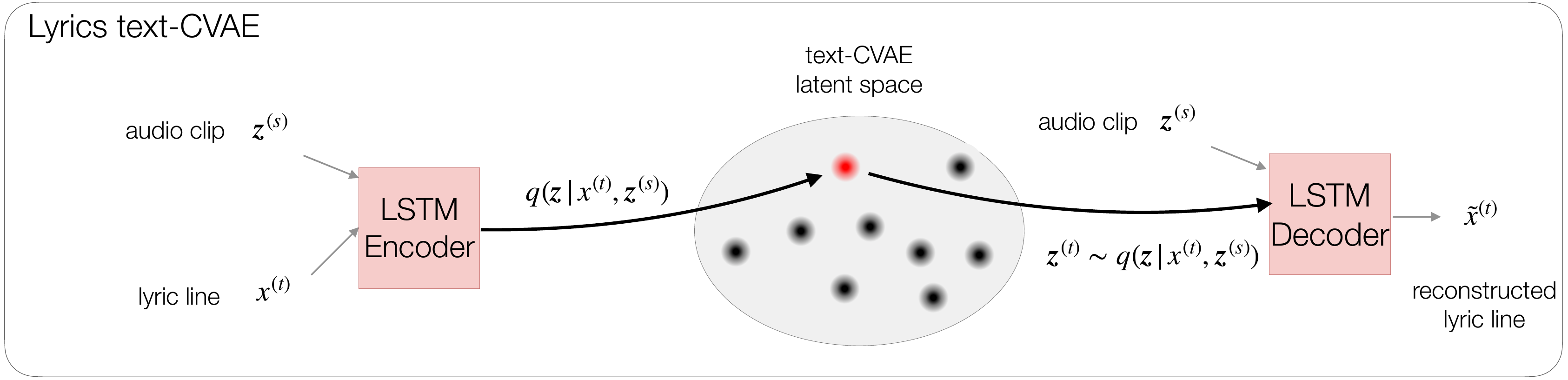}
	\caption{Training text-CVAE.}
	\label{fig:text-CVAE}
\end{figure*}

\noindent \textbf{Text generation at inference time.}
We use the trained model to generate novel lyric lines at inference time.
The encoder transforms the input sequence of words $x$ into the approximate posterior distribution $q_\phi(\bm z|x)$ learned by optimizing parameters $\phi$ of the encoder. The decoder reconstructs $x$ from the latent variable $\bm z$, sampled from $q_\phi(\bm z|x)$. Both encoder and the decoder in our work are recurrent neural networks, specifically, Long Short Term Memory networks (LSTMs)~\cite{hochreiter1997long}. 

\subsection{Training the GAN}
In this phase, we train a generative adversarial network (GAN), which predicts the latent code of the next audio clip given the latent code of the corresponding lyric line. The latent code ${\bm z}^{(t)}_{i}$ of the lyric line $x^{(t)}_{i}$ is obtained by sampling from the posterior distribution $q_{\phi}({\bm z}|x^{(t)})$ predicted by the text-CVAE.

The GAN architecture has a generator $\mathit{G}$ and a discriminator $\mathit{D}$. For training the GAN, we follow these steps:

\begin{enumerate}
	\item First, we input the lyric line $x^{(t)}_{i}$ to the encoder of the text-CVAE in order to obtain the latent code of the lyric line ${\bm z}^{(t)}_{i} = \mu^{(t)}_{i} + \tau (\epsilon \cdot \sigma^{(t)}_{i})$.
	\item After obtaining ${\bm z}^{(s)}_{i-1}$ and ${\bm z}^{(t)}_{i}$, we feed the element-wise addition of these two vectors [${\bm z}^{(s)}_{i-1}$ $\oplus$ ${\bm z}^{(t)}_{i}$] to the generator network $\mathit{G}$, which outputs a predicted latent code of the next audio clip $\hat{{\bm z}}^{(s)}_{i}$. As an alternative to element-wise addition, we also evaluated (a) element-wise multiplication (Hadamard product)  [${\bm z}^{(s)}_{i-1}$ $\circ$ ${\bm z}^{(t)}_{i}$] and (b) weighted element-wise addition [${\bm W_s}{\bm z}^{(s)}_{i-1}$ $\oplus$ ${\bm W_t}$ ${\bm z}^{(t)}_{i}$], where ${\bm W_s}$ and ${\bm W_t}$ are trainable weights matrices learned concurrently with the GAN training routine.
	\item The role of the GAN discriminator network $\mathit{D}$ is to tell apart the ``real'' data from the ``generated'' data. The element-wise addition of the generator's output $\hat{{\bm z}}^{(s)}_{i}$ with the lyric line latent code ${\bm z}^{(t)}_{i}$ is the ``generated'' data $\hat{{\bm z}}$ = [${\bm z}^{(t)}_{i}$ $\oplus$ $\hat{{\bm z}}^{(s)}_{i}$]. Whereas the ``real'' data ${\bm z}$ = [${\bm z}^{(t)}_{i}$ $\oplus$ ${\bm z}^{(s)}_{i}$] is the element-wise addition of the lyric line latent code and the actual latent code ${\bm z}^{(s)}_{i}$ of the next audio clip $x^{(s)}_{i}$, which is obtained from the spec-VAE by following the same process as in step 1.
	The discriminator $\mathit{D}$ tries to distinguish between the two types of inputs (``real'' vs. ``generated''). This adversarial training regime incentivizes the generator $\mathit{G}$ to match $\hat{{\bm z}}^{(s)}$ as closely as possible to ${\bm z}^{(s)}$.
\end{enumerate}

The adversarial loss is formulated as follows:
\begin{equation}
\begin{split}
    \min_G \max_D V(D, G) &= \mathbb{E}_{x \sim \mathcal{D}_\text{train}}[\textrm{log} D({\bm z}^{(s)}) + \textrm{log}(1 - D(\hat{\bm z}^{(s)})]
\end{split}
\end{equation}
where $\mathcal{D}_\text{train}$ is the training data, and each sample $x=\{x^{(s)}_{i-1}, x^{(t)}_{i}, x^{(s)}_{i}\}$.
We also add an auxiliary MSE loss to the objective function as it is found to stabilize GAN training~\cite{khan-etal-2020-adversarial}.
The overall loss for the GAN is:
\begin{equation}
    J_{GAN} = \min_G \max_D V(D, G) + \lambda_{MSE} || \hat{{\bm z}}^{(s)} - {\bm z}^{(s)} ||
\end{equation}
\begin{figure*}[!t]
	\centering \small
	\includegraphics[width=1\linewidth]{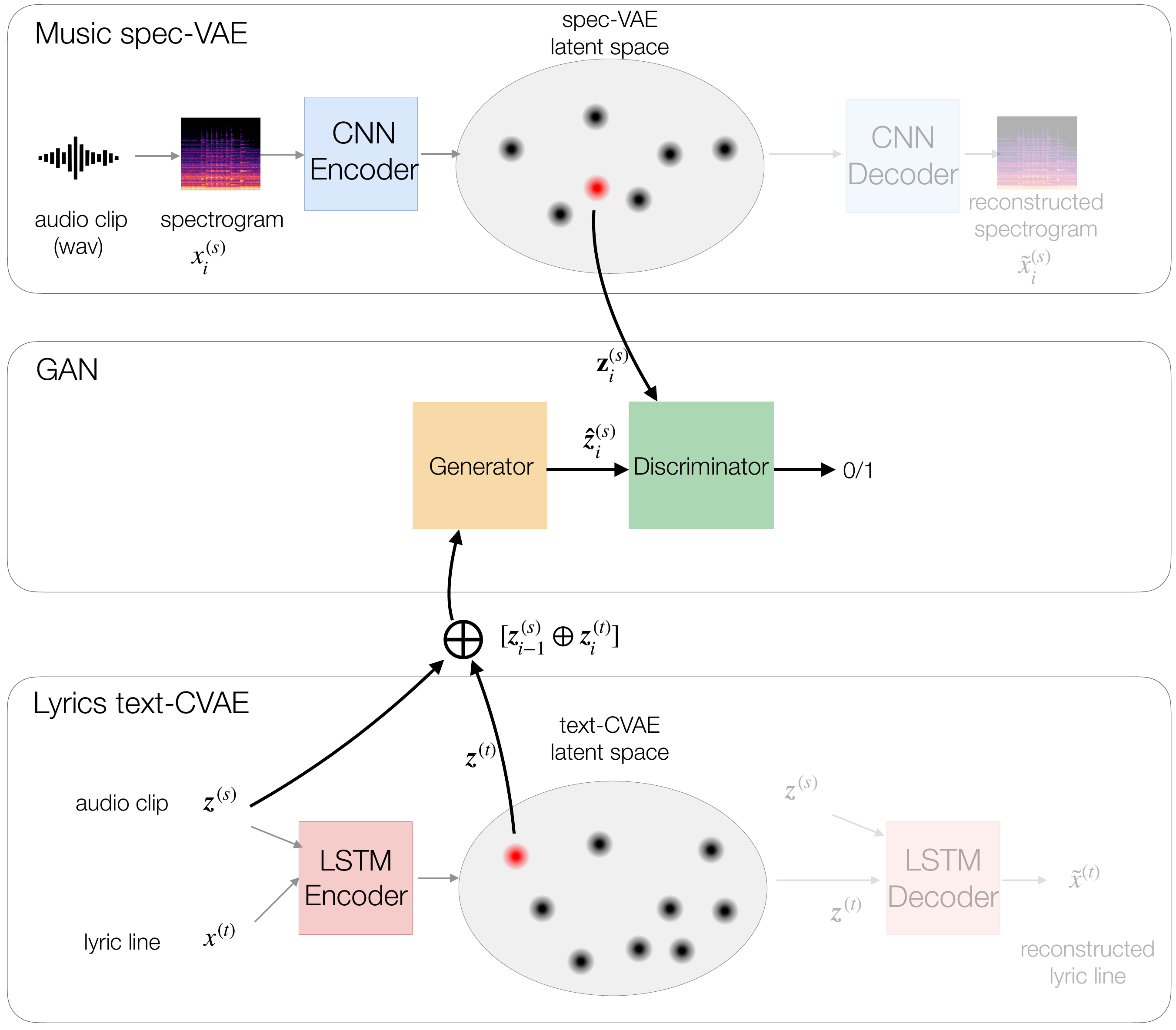}
	\caption{Training GAN.}
	\label{fig:training-GAN}
\end{figure*}

\subsection{Retrieval module}
The system requires a collection of music audio clips $X^{(s)}_{data}$ that the model would draw from to dynamically create a generative music audio stream. According to the intention of the application, this dataset could be comprised of the recordings from the artist's studio sessions. After the spec-VAE has been trained, we run it in inference mode by feeding each audio clip $x^{(s)} \in X^{(s)}_{data}$ to obtain the latent code ${\bm z}^{(s)}$ by sampling from the posterior distribution $q_{\phi}({\bm z}|x^{(s)})$. This results in a set ${\bm Z}^{(s)}$ of all latent codes.

After the GAN generator $\mathit{G}$ outputs the predicted latent code $\hat{{\bm z}}^{(s)}_i$ of the next audio clip, the latent code is sent to the retrieval module, which uses cosine similarity to rank all latent codes ${\bm z}^{(s)} \in {\bm Z}^{(s)}_{data}$ of the audio clips in the music collection $X^{(s)}_{data}$. The latent code ${\bm z}^{(s)}$ of the next audio clip is selected using either argmax or top-$K$ sampling, where $K$ could be a user-controlled hyperparameter. The audio clip $x^{(s)}_{i}$ corresponding to the selected latent code is then added to the generative music stream, played to the user of the application.

\subsection{Inference}
\begin{figure*}[!t]
	\centering \small
	\includegraphics[width=1\linewidth]{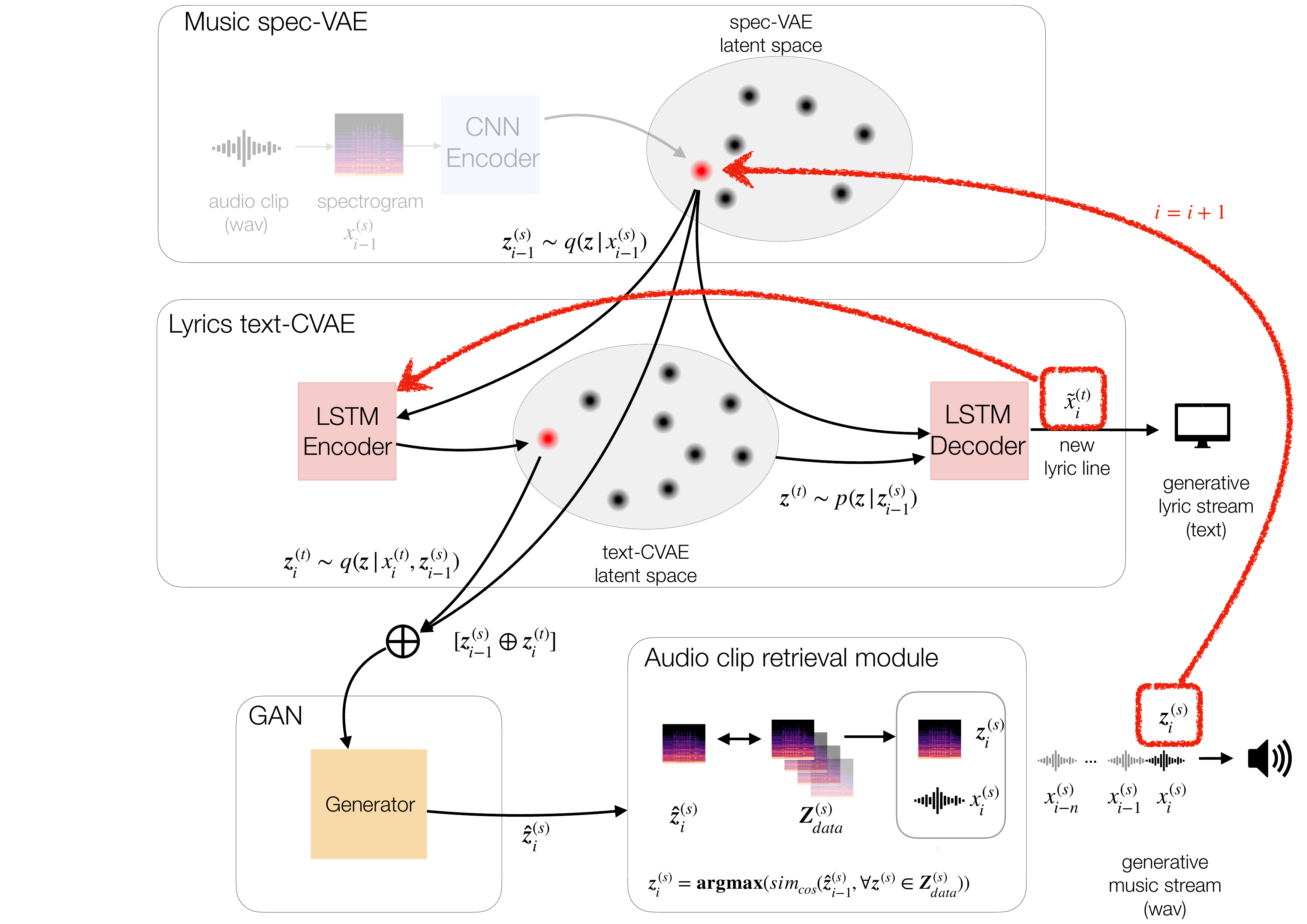}
	\caption{Inference without user input (autonomous mode).}
	\label{fig:inference-loop}
\end{figure*}

The system can run indefinitely in a fully autonomous mode without user input. In this mode the generated lyric line and currently playing audio clip influence the prediction of the next audio clip, which in turn influences the generation of the next lyric line, and so on. In more detail, the process (Figure~\ref{fig:inference-loop}) consists of the following steps.

\begin{enumerate}
    \item The process is initially seeded with a random audio clip $x^{(s)}_{i-i}$ from the user's collection $X^{(s)}_{data}$. Its corresponding latent code ${\bm z}^{(s)}_{i-i}$ is obtained by sampling from the posterior distribution predicted by spec-CVAE.
    \item The latent code ${\bm z}^{(s)}_{i-i}$ is then sent to the decoder of the text-VAE, which concatenates it with the lyric latent code ${\bm z}^{(t)}$ sampled from the text-CVAE prior distribution. The decoder generates a batch of 100 lyric lines.
    \item The generated lyric lines are then ranked by BERT fine-tuned on a custom dataset of high/low quality generated lyric lines \cite{vechtomovalyricjam}. A lyric line $x^{(t)}_{i}$ selected through top-K sampling (K=10 in our experiments) is then displayed to the user.
    \item The generated lyric line is then fed to the text-CVAE encoder to obtain its latent code ${\bm z}^{(t)}_{i}$
    \item The element-wise addition of the latent codes of the lyric line and of the previous audio clip [${\bm z}^{(s)}_{i-1}$ $\oplus$ ${\bm z}^{(t)}_{i}$] is then fed to the GAN generator $\mathit{G}$, which predicts the latent code of the next audio clip $\hat{{\bm z}}^{(s)}_{i}$.
    \item The latent code output by the generator is then fed to the retrieval module where it is matched to all latent codes ${\bm Z}^{(s)}_{data}$ in the user's collection, returning the audio clip $x^{(s)}_{i}$ that is added to the generative audio stream played to the user.
    \item The audio clip played to the user is then used to initialize the next iteration of the autonomous mode loop (step 1).
\end{enumerate}

\begin{figure*}[!t]
	\centering \small
	\includegraphics[width=1\linewidth]{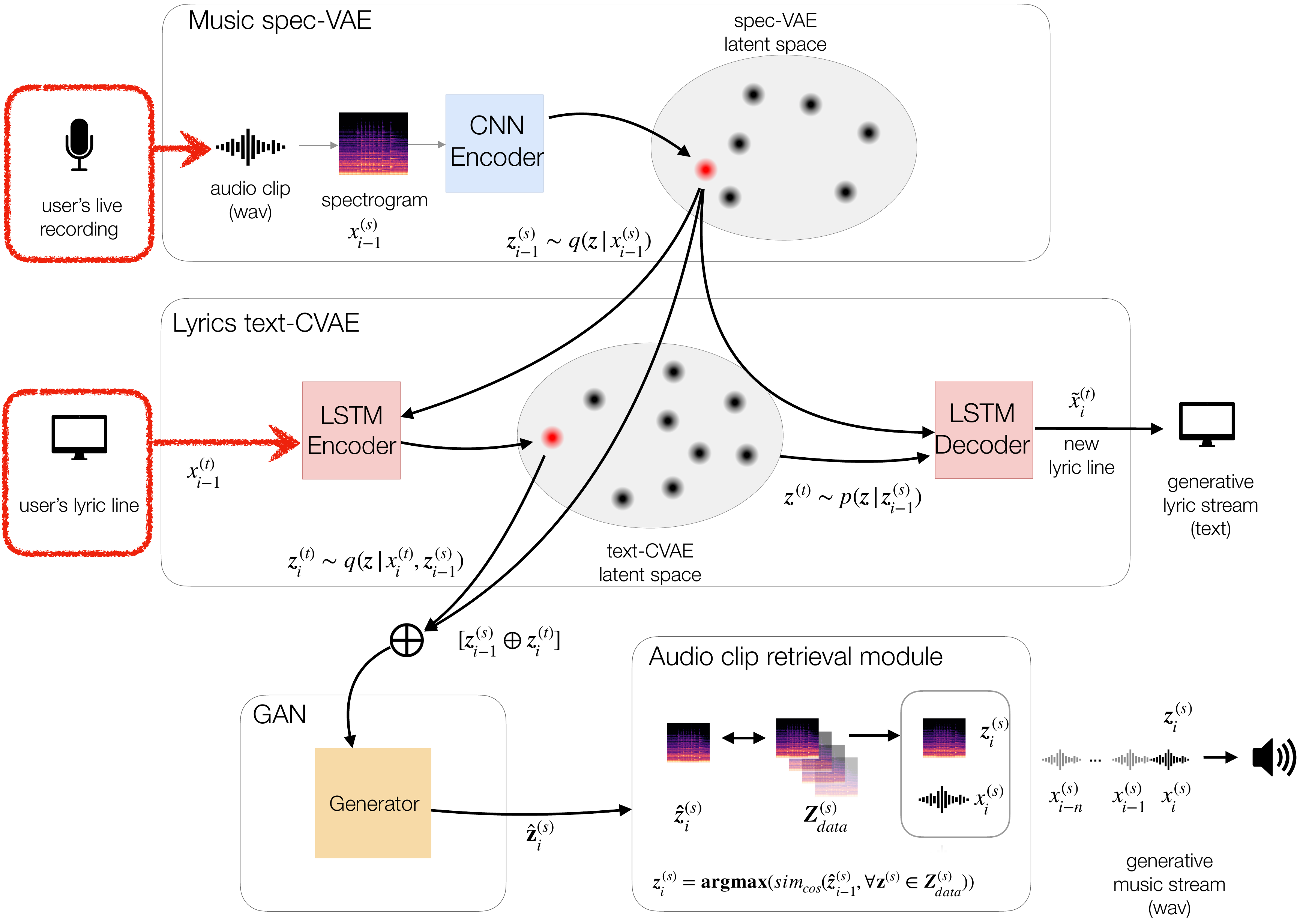}
	\caption{Inference with optional user input.}
	\label{fig:inference-user}
\end{figure*}

At any time, the system can accept user's input in the form of either a live-recorded audio clip or a new lyric line or both (Figure~\ref{fig:inference-user}). This is intended as a mechanism for the user to influence the generative process, and steer the system in a certain direction of user's choosing. For example, if the system is playing ambient music, the user can play a heavy drum or bass track to steer the generative process into that genre. If the user records an audio clip, it replaces the $x^{(s)}_{i}$ in Step 7 above, and is used to initialize the subsequent iteration through the system. If the user provides a lyric line, it interrupts the feedback loop in Step 4, replacing the system-generated lyric line. After one iteration, the system returns to using its own generated lyric lines unless the user provides a new line.

\section{Data and evaluation}
\subsection{Data}
\noindent \textbf{Spec-VAE.}
For training the Spec-VAE, we combine clips from two sources:

\begin{enumerate}
    \item A lyric-music aligned dataset~\cite{vechtomovaLyrics2020}, which contains 22,763 WAV audio clips of original songs by seven music artists in various sub-genres of ``Rock'', including two artists who create Electronic rock sub-genre with extensive use of synthesizers.
    \item Our newly created dataset of 4814 10-second WAV audio clips (over 13 hours) of original electronic instrumental music created by an electronic music artist during studio recording sessions over a period of one year. We will refer to our newly created dataset as $X^{(s)}_{data}$. Over 20 different electronic and electroacoustic instruments were used in the recording of these sessions, which can be grouped into the following categories: drone synthesizers (2249 clips), percussive synthesizers (960), keyboard synthesizers and piano (2095), electric guitar (386). Many of the tracks contain more than one category of instruments, hence the sum of the number of clips shown in brackets is greater than the total number of clips.
\end{enumerate}

\noindent \textbf{Text-CVAE and GAN.}
To train the Text-CVAE, we use the \textit{aligned} lyric-music dataset in Vechtomova et al.~\cite{vechtomovaLyrics2020}, which contains 18,000 lyric lines and their corresponding audio clips.

To train the GAN, we encode the clips and their corresponding lyric lines using trained Spec-VAE and Text-CVAE, respectively, to obtain a dataset of latent codes.
We prepare the latent codes beforehand to reduce GAN's training time.




\subsection{Implementation Details}
All models were trained using one NVIDIA GeForce RTX 2080 Ti with a VRAM of 12G.
\subsubsection{Spec-VAE.}
We implement the Spec-VAE in PyTorch~\cite{Pytorch2019}.
The encoder has four Conv2d layers interleaved with ReLU activation function~\cite{nair2010rectified}.
The kernel size and stride of all the Conv2d layers is set to 6 and 2, respectively.
The decoder consists of four ConvTranspose2d layers, interleaved with three ReLU activations and one Sigmoid activation in the end.
We use 128 latent dimensions to express the mean and sigma vectors.
We train the Spec-VAE on for 200 epochs with a batch size of 32, learning rate of 1e-4, 0.2 dropout, and Adam optimizer~\cite{kingma2014adam}.
We set the sampling temperature to 1.0 for both training and inference.

\subsubsection{Text-CVAE.}
We use Tensorflow~\cite{abadi2016tensorflow} to implement the Text-CVAE.
We use a single-layer bi-LSTM encoder and an LSTM decoder with 300 embedding dimensions, and 128 latent dimensions to represent the encoded text.
We train the Text-CVAE for 200 epochs with batch size of 100, and a word dropout probability of 0.2.
We anneal the learning rate from a value of 5e-3 to 1e-5 over 3000 steps.

\subsubsection{GAN.} 
We use the AllenNLP library~\cite{gardner2018allennlp} to implement the GAN.
The generator and discriminator networks are 3-layered feed-forward neural networks, interleaved with ReLU activation function.
We train the GAN alignment network for 20 epochs with a batch size of 32, a learning rate of 1e-3, and Adam optimizer for both the generator and the discriminator.
We set $\lambda_{MSE}=2.0$ and sampling temperature to 1.0 during both training and inference.

\subsubsection{BERT.}
We use the Transformers library~\cite{wolf-etal-2020-transformers} to fine-tune a BERT-base model for sequence classification on our custom dataset.
We fine-tune the model for 15 epochs with a batch size of 16, using a learning rate warmup scheduler for the first 500 steps of training, with a weight decay of 0.01.

\subsection{Evaluation}

Our first goal is to evaluate the effectiveness of the conditional GAN in predicting subsequent audio clips that are congruent with the previous clip. Specifically, we set out to determine the percentage of the clips from the same composition among the top-ranked clips with respect to the previous clip. We ran the system in autonomous mode for one hour. In each iteration, we selected the conditioning clip randomly from the entire collection, which was then used to generate a lyric line. Both of their latent codes were then used as inputs to the GAN module to predict the latent code of the next clip, which was used by the Retrieval module to rank clips in the entire collection. We recorded top 50 clips ranked by the Retrieval module, and calculated Precision at 50, 20, 10, 5 and 1. The results are summarized in Table~\ref{tab:precision-results}. We also analyzed precision by types of conditioning clips based on the category of the lead instrument in the clip: (a) Drone, (b) Piano and (c) Rhythm/Percussion. 

\begin{table}
\caption{Precision values at different cutoff points of the clips ranked by the Retrieval module in response to the GAN-predicted latent code generated based on the previous clip and lyric line. In calculating the precision values, a retrieved clip is considered ``relevant'' if it comes from the same composition as the previous (conditioning) clip. \label{tab:precision-results}}
\begin{tabular}{|l|l|l|l|l|l|}
\hline
\textbf{Conditioning clips}	& \textbf{P@50}	& \textbf{P@20} & \textbf{P@10} & \textbf{P@5} & \textbf{P@1} \\ 
\hline
Drone (46 clips)	        & 0.3083   &   0.3667   &   0.4087  & 0.4783 & 0.5217 \\
Piano (36 clips)	        & 0.2111	& 0.2600   &   0.2750   &   0.3000  & 0.3889 \\
Rhythm/Percussion (22 clips)	    & 0.2336	& 0.3000   &   0.3363   &   0.4091  & 0.3636 \\
\hline
All instruments	(104 clips)	    & 0.2588	& 0.3157   &   0.3471   &   0.4019  & 0.4423 \\
\hline
\end{tabular}
\end{table}

\begin{figure*}
	\centering \small
	\includegraphics[width=0.7\linewidth]{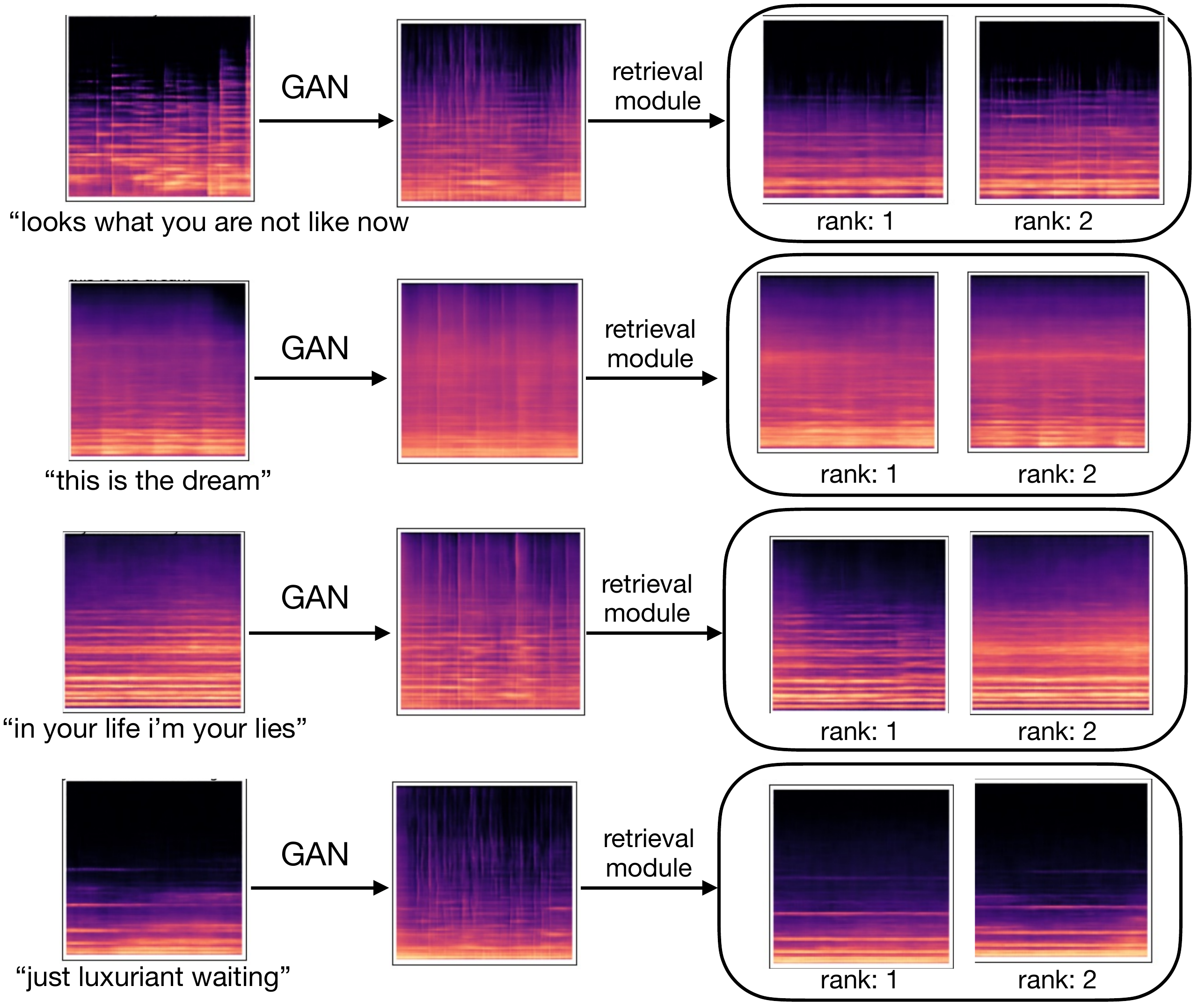}
	\caption{Conditional GAN spectrogram predictions based on the previous clip and lyric lines. Each row shows the spectrogram of the previous clip with a lyric line generated from it, the generated spectrogram from the GAN-predicted latent code and spectrograms of the top two clips returned by the retrieval module. }
	\label{fig:gan-example-specs}
\end{figure*}

Figure~\ref{fig:gan-example-specs} shows four sets of spectrograms, where the left-most figure in each row is the previous clip's spectrogram with a lyric line generated from it. The latent codes of this spectrogram and the lyric line are used as the input to the GAN. The spectrogram in the second image in each row is generated based on the latent code predicted by the GAN. The last two spectrograms in each row are the top first and second spectrograms returned by the Retrieval module. The Figure illustrates visual differences in the GAN-predicted spectrograms based on different previous clip and lyric line combinations, as well as differences in the corresponding top-ranked retrieved clips. The spectrogram generated from the GAN-predicted latent code bears overall visual resemblance to its conditioning spectrogram, and both have similarities with the top retrieved spectrograms. 

The results show that quite a substantial number of top-ranked clips come from the same composition. This is especially the case in the Drone category, where in 52\% of cases the top-ranked clip (P@1) comes from the same composition as the conditioning clip. This points to the good ability of the system to predict clips that are musically related to the previous clip, resulting in a coherent music. However, the fact that not all top-ranked clips are from the same composition is essential, as this means that the music can evolve and not be stuck in the same composition.

\begin{figure*}
	\centering \small
	\includegraphics[width=0.4\linewidth]{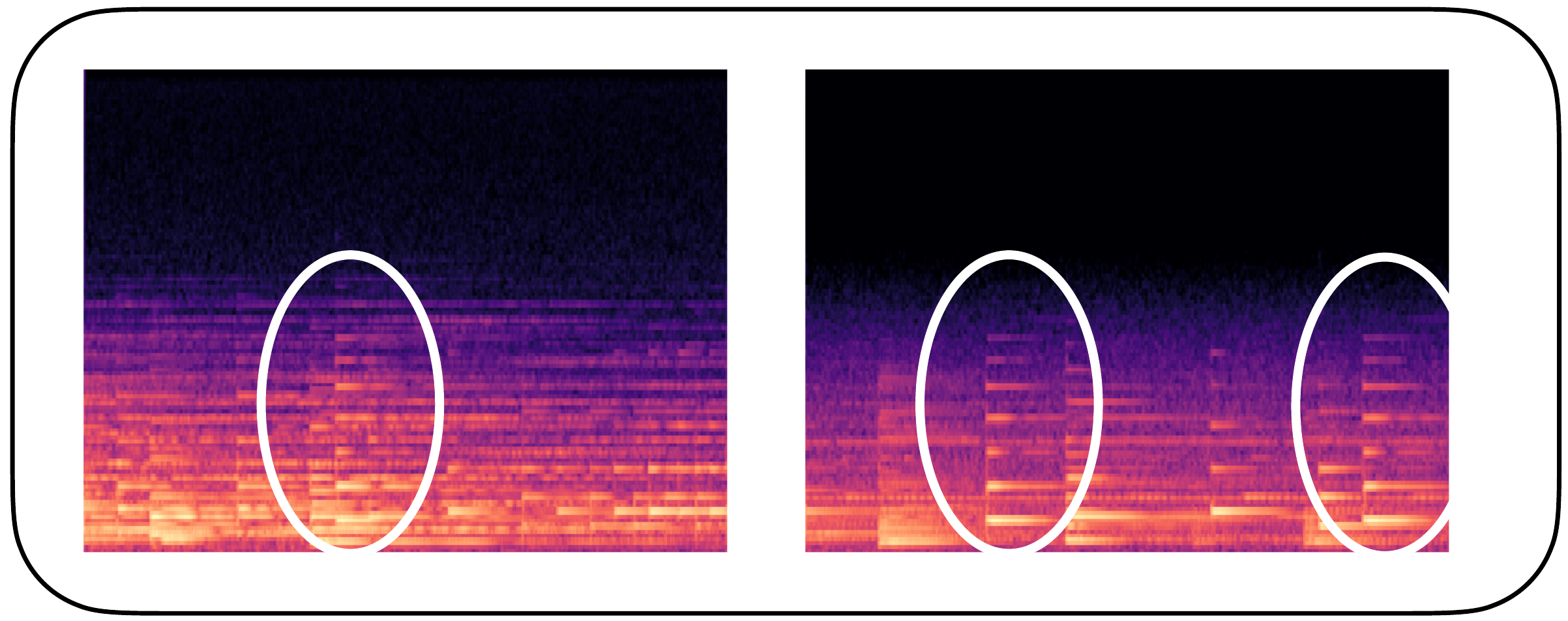}
	\caption{Example of the two piano clips from different compositions in the same key. The left-hand spectrogram corresponds to the previous clip, while the right-hand spectrogram represents one of the top-ranked clips (rank 13). The matching chords are highlighted. }
	\label{fig:piano-specs}
\end{figure*}

One research question we would like to explore is how sensitive the system is in predicting related piano compositions. This is an interesting question, because most piano compositions in our dataset have a motif and are typically played in a specific key, so prediction of the next clip from a composition in a different key may clash with the previous clip and not result in a smooth transition or an aesthetically pleasing music. For the 36 piano clips in Table~\ref{tab:precision-results}, the total number of predicted clips among the top 10 that come from the same piano composition as the conditioning clip is 99, while the number of clips from a different piano composition is only 10. This suggests that the model is able to predict piano clips that come from the same composition and therefore have similar musical characteristics, such as key. Furthermore, upon listening to the 10 clips that came from a different composition, we observed that they evoke a similar mood to the conditioning clip and are typically played in the same key. Figure~\ref{fig:piano-specs} demonstrates visual similarities between the spectrograms of two such clips. Specifically, yellow horizontal lines at different heights in the spectrograms correspond to the frequencies of the played notes or chords. This provides further evidence that the GAN is able to generate a latent code representing a clip that is close to the preceding clip at a fine level of granularity, including the notes or chords played.

\begin{figure*}
	\centering \small
	\includegraphics[width=0.7\linewidth]{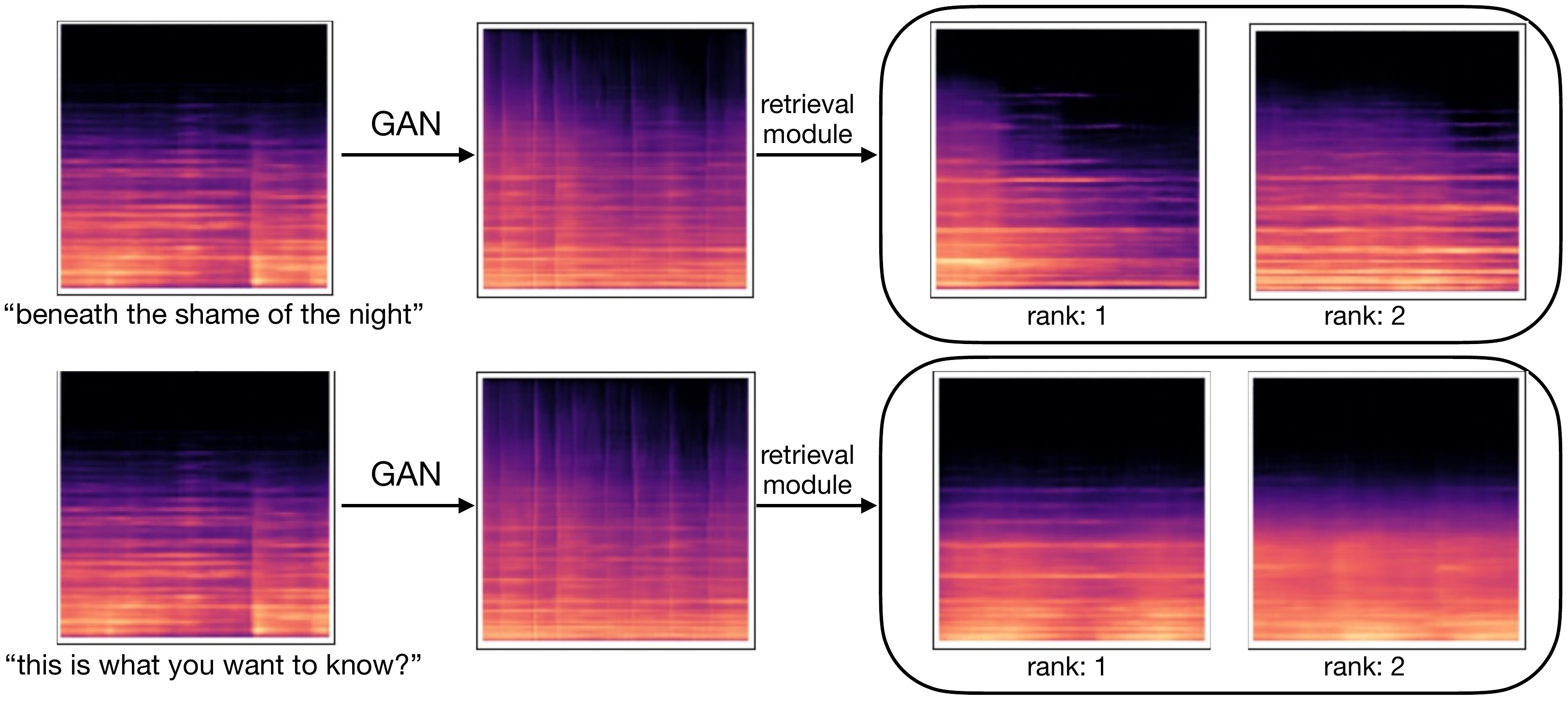}
	\caption{Conditional GAN spectrogram predictions based on the previous clip and lyric lines. To demonstrate the effect of the lyric line on the prediction of the next clip, the previous audio clip is fixed, while the lyric line is different.}
	\label{fig:gan-example-specs-fixed-conditioning-spec}
\end{figure*}

To analyze the extent that GAN predictions are influenced by lyric lines, we fixed the conditioning spectrogram and varied the lyric lines. Figure~\ref{fig:gan-example-specs-fixed-conditioning-spec} shows examples from the system with this setting. The lyric line has an effect on the predicted spectrogram, as can be seen both from the visual differences in the GAN-predicted spectrograms (the second image in each row) and in the differences of the top two retrieved clips (last two images in each row).

To quantitatively evaluate the effect of the lyric line on the next clip prediction, we randomly selected seven clips in each of the three categories: Drone, Piano, Rhythm. We let the system run for 10 iterations in autonomous mode with a given clip fixed as the conditioning clip. In each iteration, the system randomly selected a lyric line from the top 10 lyric lines generated by text-CVAE conditioned on this clip and ranked by the fine-tuned BERT model. Thus, for each conditioning clip we obtained set $N$, which is a union of $n$ top-ranked clips across all 10 iterations. We then computed Impact@$n$ as the number of unique clips in set $N$ out of the total number of clips in $N$. In our experiments, $n$ was set to 10, 5 and 2. Table~\ref{tab:diversity-results} shows the results of the evaluation per category and for all clips.

\begin{table}
\caption{Impact of variable lyric lines on the top-ranked clips, given a fixed conditioning clip. Impact is measured at 10, 5 and 2 top-ranked clips.  \label{tab:diversity-results}}
\begin{tabular}{|l|l|l|l|}
\hline
\textbf{Conditioning clips}	& \textbf{Impact@10}	& \textbf{Impact@5} & \textbf{Impact@2} \\ 
\hline
Drone (7 clips)	        & 0.1443   &   0.1771  & 0.1929 \\
Piano (7 clips)	        & 0.1943	& 0.2171   &   0.2357 \\
Rhythm/Percussion (7 clips)	    & 0.1471	& 0.1686   &   0.1857  \\
\hline
All instruments	(21 clips)	    & 0.1619	& 0.1876   &   0.2048  \\
\hline
\end{tabular}
\end{table}

The results in the Table show that on average 20.48\% of all clips among the top-2 clips predicted for the same previous clip are unique, which indicates that the lyric line has a noticeable effect on the prediction of the next clip. The effect of the previous clip on the prediction is stronger than the effect of the lyric line, which is desirable, as we would like the next clip to be musically consistent with the previous clip. The lyric line therefore acts more as a nudge towards a certain mood or a musical theme, rather than causing a drastic change in the generated music. The variance calculated on the 21 clips is 0.0019, 0.0036 and 0.0045 for the top 10, 5 and 2 sets, respectively. While the variance is low, we observed that some conditioning clips have lower diversity than others. When a clip comes from a composition that is highly dissimilar to the others in the collection, the top-ranked clips are more likely to also come from the same composition. This is because the variations attributed to the lyric line may not be strong enough for the GAN to generate a sufficiently different latent code that would make it close to clips from the other compositions. 

\section{Listening tests}
We conducted a listening test study to determine whether the system constructs more musically coherent compositions compared to audio segments made of randomly selected audio clips. We recorded 30 1-minute audio segments by running \methodname in autonomous mode (test condition). Similarly, we recorded 30 1-minute audio segments by letting the system pick random clips in each iteration (control condition). The segments in control and test conditions were randomly paired, and were given to the participants. The order of test and control conditions in each pair was randomly assigned.

Two participants familiar with electronic music genre participated in the study. Each participant was given the same 30 pairs of audio segments. For each pair, the participant was asked to listen to each audio segment, and select the one which they perceived to be more musically coherent. As a result of the study, 22 and 25 of the test condition clips were determined by Participants A and B, respectively, to be more coherent than the control condition clips. The overall accuracy is 78.3\%. This suggests that the compositions created by \methodname are perceived to be more musically coherent than segments with randomly selected clips.

\section{Interactive application}
We developed a real-time web-based interactive application in ReactJS and NodeJS, with the AI backend written in Python. Once the user starts the application, it goes into the autonomous mode by default, playing continuous music from the audio clips predicted in real time based on the lyric lines, which are also generated in real-time based on the previously played audio clip. The two modalities thus feed into each other: the next audio clip is predicted by the lyric line and previous clip, and is in turn used to condition the generation of the next lyric line. The lyric lines are continuously displayed as animated text that appears, slowly drifts away and fades into the background. The audio clips predicted by the system are added to the music using a crossfade. To create aesthetically pleasing music, the system randomly varies the duration of the clips from 10 to 40 seconds. The duration of crossfade is also variable within a predetermined range.

\begin{figure*}
	\centering \small
	\includegraphics[width=0.8\linewidth]{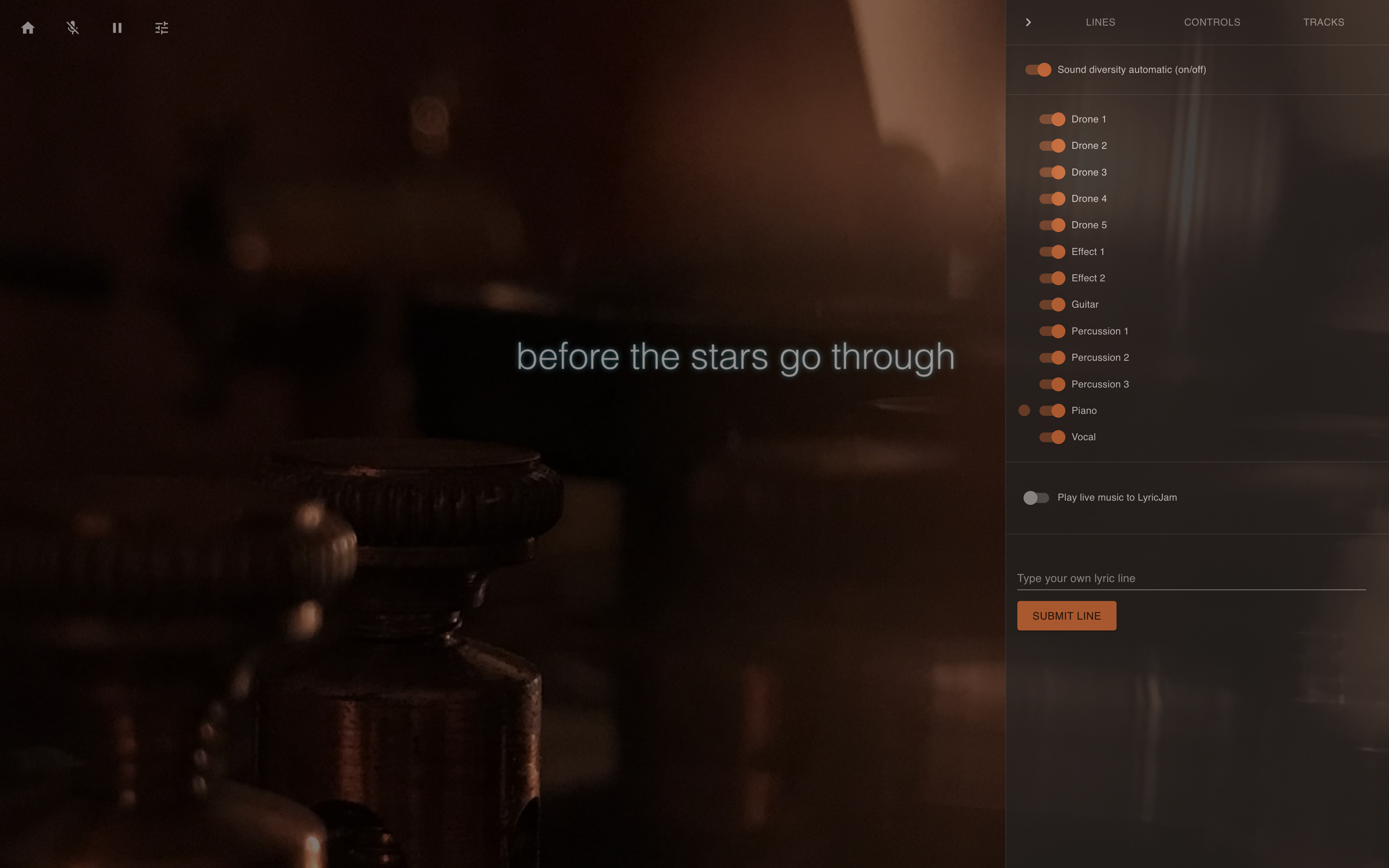}
	\caption{\methodname\space-- interactive web-based application developed in ReactJS and NodeJS with the AI backend, written in Python. The app plays generative music and displays corresponding generated lyrics. The expandable panel on the right has three tabs, showing the history of generated lines and played clips, as well as controls that allow the user to (a) adjust sound diversity or set it to automatic, (b) include/exclude audio clips with specific instruments (c) go to the live performance mode where the system predicts the subsequent clips and generates lyrics based on the music played by the user (d) type their own lyric line that will influence the next audio clip prediction }
	\label{fig:interface-controls}
\end{figure*}

\subsection{Autonomous mode}
 By default, the system controls the diversity of the audio clips retrieved by the system. The diversity value $k$ is the number of top clips ranked by the cosine similarity of their ${\bm z}^{(s)}$ vector with respect to the GAN-predicted $\hat{{\bm z}}^{(s)}$ vector. We implemented Perlin noise algorithm \cite{perlin}, which is a type of gradient noise originally designed to create natural appearing textures (e.g. waves) in computer graphics. Perlin noise in \methodname is used to determine variable $k$, enabling the generative music to contain passages of more closely related clips when $k$ value is low, periodically alternated with more dramatic shifts in sounds when $k$ is high. Users can also choose to control the diversity value $k$ manually to more precisely suit their creative goals. Thus, they can narrow down the predicted clip to the one which most closely matches the previous clip, or explore more distant clips and get inspiration from potentially unexpected musical combinations of sounds. 

Each music composition in our dataset is annotated with the list of music instruments. As each audio clip is playing, the system indicates which instruments are being played. By default, all instruments are selected, which means that any clip from the collection could be potentially included in the music. If a user deselects a given instrument, all clips containing this instrument will be excluded by the Ranking module. This gives  users a more fine-grained mechanism to control what instruments they want to hear. 

The system also lets the user select any of the past played clips as the next clip to be played, overriding the system-predicted clip. The user can also pin each audio clip, thus forcing that clip to be the conditioning clip for both lyric generation and subsequent audio clip prediction. In addition, users can type their own lyric lines, which override the system-generated lyric lines, and are used by the system in the prediction of the next audio clip.

\subsection{Live performance mode}
In live performance mode, the system ``listens'' to the user's audio source (microphone or an audio interface device), recording 10-second clips at short intervals. Each user-recorded audio clip is converted into a spectrogram, which is then transformed by Spec-VAE into the latent code ${\bm z}^{(s)}$. It is used to override the system-predicted clip, and is used to condition the lyric generation and prediction of the next clip. When this mode is enabled, the user can jam with the system by playing live instruments, while the system creates in real-time a musical composition and lyrics from the clips it predicts based on the user's audio input. This mode may be used by the musician to find clips that are similar to or that go well with the music they are playing, or to let the system create a continuous accompaniment to their performance. Coupled with the functionality that allows the user to include/exclude specific instruments, this mode gives the user control over the type of accompaniment they want the system to create. This feature can be used by musicians to rediscover compositions from their own catalogues and try them out as they are developing their current musical ideas.

\subsection{Collecting user feedback for model retraining}
Users can select lines that they like, which are logged on the server together with the clip that they were used to predict. Periodically, once sufficient number of lines has been collected, they would act as an augmented dataset to be added to the original dataset to retrain the system. The goal is to gather additional data that may help the system continuously learn better associations between lyrics and music, as well as learn a better lyric generation model. 

\section{Conclusions}
We presented \methodname, a real-time system that dynamically creates a continuous stream of music from the clips in the user's own catalogue of music recordings and generates corresponding lyrics. The self-perpetuating system can work in autonomous mode, whereby a generated lyric line and a previously played audio clip are used to predict the next audio clip, which in turn conditions the generation of the next lyric line. The system can also be used in live performance mode, whereby the musician plays live music, and the system ``listens'' and creates a continuous stream of music and lyrics in response to live music. The user can also influence the system by providing their own lyric lines. 

The system consists of a spectrogram variational autoencoder, which is trained on a collection of audio clips, a conditional text variational autoencoder, which is trained to generate lyric lines conditioned on the latent code of the corresponding audio clip, encoded by the spec-VAE. The system also contains a generative adversarial network that predicts the latent code of the next audio clip based on the latent codes of the previously generated lyric line and the preceding audio clip. A retrieval module then retrieves an audio clip, whose latent code closely matches the GAN-predicted latent code by the cosine similarity measure. An interactive web-based application has been developed that creates generative music compositions and corresponding lyrics in real-time. 

The results of a listening study showed that participants perceived the compositions generated by \methodname to be more musically coherent than audio segments composed of random audio clips. Furthermore, automatic evaluation demonstrated that while the top ranked clips contain many clips from the same composition, they also include clips from other compositions. This suggests that while the generated musical compositions have musical coherence, they also have musical diversity, which allow the composition to musically develop over time. Finally, the results of evaluation demonstrated a noticeable impact of generated lyric lines on the selection of the following audio clips.

\bibliographystyle{unsrt}  
\bibliography{references}

\begin{thebibliography}{10}

\bibitem{vechtomovaLyrics2020}
Olga Vechtomova, Gaurav Sahu, and Dhruv Kumar.
\newblock Generation of lyrics lines conditioned on music audio clips.
\newblock In {\em Proceedings of the 1st Workshop on NLP for Music and Audio
  (NLP4MusA)}, pages 33--37, Online, oct 2020. Association for Computational
  Linguistics.

\bibitem{vechtomovalyricjam}
Olga Vechtomova, Gaurav Sahu, and Dhruv Kumar.
\newblock Lyricjam: A system for generating lyrics for live instrumental music.
\newblock In {\em Proceedings of the 12th Conference on Computational
  Creativity}, 2021.

\bibitem{Hunt2017ThoughtsOI}
Samuel~J. Hunt, Tom Mitchell, and Chris Nash.
\newblock Thoughts on interactive generative music composition.
\newblock 2017.

\bibitem{briot2019deep}
J.P. Briot, G.~Hadjeres, and F.D. Pachet.
\newblock {\em Deep Learning Techniques for Music Generation}.
\newblock Computational Synthesis and Creative Systems. Springer International
  Publishing, 2019.

\bibitem{oord2016wavenet}
Aaron van~den Oord, Sander Dieleman, Heiga Zen, Karen Simonyan, Oriol Vinyals,
  Alex Graves, Nal Kalchbrenner, Andrew Senior, and Koray Kavukcuoglu.
\newblock Wavenet: A generative model for raw audio.
\newblock {\em arXiv preprint arXiv:1609.03499}, 2016.

\bibitem{DUA2020465}
Mohit Dua, Rohit Yadav, Divya Mamgai, and Sonali Brodiya.
\newblock An improved rnn-lstm based novel approach for sheet music generation.
\newblock {\em Procedia Computer Science}, 171:465--474, 2020.
\newblock Third International Conference on Computing and Network
  Communications (CoCoNet'19).

\bibitem{lstmmusicgeneration}
Sarthak Agarwal, Vaibhav Saxena, Vaibhav Singal, and Swati Aggarwal.
\newblock Lstm based music generation with dataset preprocessing and
  reconstruction techniques.
\newblock In {\em 2018 IEEE Symposium Series on Computational Intelligence
  (SSCI)}, pages 455--462, 2018.

\bibitem{conditional_drums}
Dimos Makris, Guo Zixun, Maximos Kaliakatsos-Papakostas, and Dorien Herremans.
\newblock Conditional drums generation using compound word representations.
\newblock In {\em Artificial Intelligence in Music, Sound, Art and Design: 11th
  International Conference, EvoMUSART 2022, Held as Part of EvoStar 2022,
  Madrid, Spain, April 20–22, 2022, Proceedings}, page 179–194, Berlin,
  Heidelberg, 2022. Springer-Verlag.

\bibitem{BrianEno}
Brian Eno.
\newblock Generative music.
\newblock \url{http://www.inmotionmagazine.com/eno1.html}, 1996.
\newblock Accessed: 2022-10-17.

\bibitem{eigenfeldt2011negotiated}
Arne Eigenfeldt and Philippe Pasquier.
\newblock Negotiated content: Generative soundscape composition by autonomous
  musical agents in coming together: Freesound.
\newblock In {\em ICCC}, pages 27--32, 2011.

\bibitem{thorogood2012audio}
Miles Thorogood, Philippe Pasquier, and Arne Eigenfeldt.
\newblock Audio metaphor: Audio information retrieval for soundscape
  composition.
\newblock {\em Proc. of the Sound and Music Computing Cong.(SMC)}, pages
  277--283, 2012.

\bibitem{Turchet2020VoicebasedIF}
Luca Turchet and Alex Zanetti.
\newblock Voice-based interface for accessible soundscape composition:
  composing soundscapes by vocally querying online sounds repositories.
\newblock {\em Proceedings of the 15th International Audio Mostly Conference},
  2020.

\bibitem{seqgan-melodyconditioned}
Yihao Chen and Alexander Lerch.
\newblock Melody-conditioned lyrics generation with seqgans.
\newblock In {\em 2020 IEEE International Symposium on Multimedia (ISM)}, pages
  189--196, 2020.

\bibitem{malmi2016dopelearning}
Eric Malmi, Pyry Takala, Hannu Toivonen, Tapani Raiko, and Aristides Gionis.
\newblock Dopelearning: A computational approach to rap lyrics generation.
\newblock In {\em Proceedings of the 22nd ACM SIGKDD International Conference
  on Knowledge Discovery and Data Mining}, pages 195--204, 2016.

\bibitem{oliveira2015tra}
Hugo~Gon{\c{c}}alo Oliveira.
\newblock Tra-la-lyrics 2.0: Automatic generation of song lyrics on a semantic
  domain.
\newblock {\em Journal of Artificial General Intelligence}, 6(1):87, 2015.

\bibitem{potash2015ghostwriter}
Peter Potash, Alexey Romanov, and Anna Rumshisky.
\newblock Ghostwriter: Using an lstm for automatic rap lyric generation.
\newblock In {\em Proceedings of the 2015 Conference on Empirical Methods in
  Natural Language Processing}, pages 1919--1924, 2015.

\bibitem{watanabe2018melody}
Kento Watanabe, Yuichiroh Matsubayashi, Satoru Fukayama, Masataka Goto, Kentaro
  Inui, and Tomoyasu Nakano.
\newblock A melody-conditioned lyrics language model.
\newblock In {\em Proceedings of the 2018 Conference of the North American
  Chapter of the Association for Computational Linguistics: Human Language
  Technologies, Volume 1 (Long Papers)}, pages 163--172, 2018.

\bibitem{kingma2014auto}
Diederik~P. Kingma and Max Welling.
\newblock {Auto-Encoding Variational Bayes}.
\newblock In {\em 2nd International Conference on Learning Representations,
  {ICLR} 2014, Banff, AB, Canada, April 14-16, 2014, Conference Track
  Proceedings}, 2014.

\bibitem{goodfellow2014generative}
Ian Goodfellow, Jean Pouget-Abadie, Mehdi Mirza, Bing Xu, David Warde-Farley,
  Sherjil Ozair, Aaron Courville, and Yoshua Bengio.
\newblock Generative adversarial nets.
\newblock {\em Advances in neural information processing systems}, 27, 2014.

\bibitem{hochreiter1997long}
Sepp Hochreiter and J{\"u}rgen Schmidhuber.
\newblock Long short-term memory.
\newblock {\em Neural computation}, 9(8):1735--1780, 1997.

\bibitem{khan-etal-2020-adversarial}
Kashif Khan, Gaurav Sahu, Vikash Balasubramanian, Lili Mou, and Olga
  Vechtomova.
\newblock Adversarial learning on the latent space for diverse dialog
  generation.
\newblock In {\em Proceedings of the 28th International Conference on
  Computational Linguistics}, pages 5026--5034, Barcelona, Spain (Online), dec
  2020. International Committee on Computational Linguistics.

\bibitem{Pytorch2019}
Adam Paszke, Sam Gross, Francisco Massa, Adam Lerer, James Bradbury, Gregory
  Chanan, Trevor Killeen, Zeming Lin, Natalia Gimelshein, Luca Antiga, Alban
  Desmaison, Andreas Kopf, Edward Yang, Zachary DeVito, Martin Raison, Alykhan
  Tejani, Sasank Chilamkurthy, Benoit Steiner, Lu~Fang, Junjie Bai, and Soumith
  Chintala.
\newblock Pytorch: An imperative style, high-performance deep learning library.
\newblock In {\em Advances in Neural Information Processing Systems 32}, pages
  8024--8035. Curran Associates, Inc., 2019.

\bibitem{nair2010rectified}
Vinod Nair and Geoffrey~E Hinton.
\newblock Rectified linear units improve restricted boltzmann machines.
\newblock In {\em Icml}, 2010.

\bibitem{kingma2014adam}
Diederik~P. Kingma and Jimmy Ba.
\newblock Adam: {A} method for stochastic optimization.
\newblock In Yoshua Bengio and Yann LeCun, editors, {\em 3rd International
  Conference on Learning Representations, {ICLR} 2015, San Diego, CA, USA, May
  7-9, 2015, Conference Track Proceedings}, 2015.

\bibitem{abadi2016tensorflow}
Mart{\'\i}n Abadi, Paul Barham, Jianmin Chen, Zhifeng Chen, Andy Davis, Jeffrey
  Dean, Matthieu Devin, Sanjay Ghemawat, Geoffrey Irving, Michael Isard, et~al.
\newblock Tensorflow: A system for large-scale machine learning.
\newblock In {\em 12th $\{$USENIX$\}$ Symposium on Operating Systems Design and
  Implementation ($\{$OSDI$\}$ 16)}, pages 265--283, 2016.

\bibitem{gardner2018allennlp}
Matt Gardner, Joel Grus, Mark Neumann, Oyvind Tafjord, Pradeep Dasigi, Nelson
  Liu, Matthew Peters, Michael Schmitz, and Luke Zettlemoyer.
\newblock Allennlp: A deep semantic natural language processing platform.
\newblock {\em arXiv preprint arXiv:1803.07640}, 2018.

\bibitem{wolf-etal-2020-transformers}
Thomas Wolf, Lysandre Debut, Victor Sanh, Julien Chaumond, Clement Delangue,
  Anthony Moi, Pierric Cistac, Tim Rault, Rémi Louf, Morgan Funtowicz, Joe
  Davison, Sam Shleifer, Patrick von Platen, Clara Ma, Yacine Jernite, Julien
  Plu, Canwen Xu, Teven~Le Scao, Sylvain Gugger, Mariama Drame, Quentin Lhoest,
  and Alexander~M. Rush.
\newblock Transformers: State-of-the-art natural language processing.
\newblock In {\em Proceedings of the 2020 Conference on Empirical Methods in
  Natural Language Processing: System Demonstrations}, pages 38--45, Online,
  oct 2020. Association for Computational Linguistics.

\bibitem{perlin}
Ken Perlin.
\newblock An image synthesizer.
\newblock In {\em Proceedings of the 12th Annual Conference on Computer
  Graphics and Interactive Techniques}, SIGGRAPH '85, page 287–296, New York,
  NY, USA, 1985. Association for Computing Machinery.

\end{thebibliography}

\end{document}